\documentclass[11pt,dvips]{article}

\usepackage{epsfig,times}
\def\Fpho{cm$^{-2}$ s$^{-1}$}
\makeatletter
\renewcommand{\section}{\@startsection{section}{1}{0in}
	{0.4\baselineskip}{0.1\baselineskip}{\Large\bf}}
\renewcommand{\subsection}{\@startsection{subsection}{2}{0in}
	{0.25\baselineskip}{-\baselineskip}{\large\bf}}
\renewcommand{\subsubsection}{\@startsection{subsubsection}{3}{0in}
	{0.1\baselineskip}{-\baselineskip}{\normalsize\bf}}
\makeatother
\begin{document}
%
%  Session and Paper Code:
\thispagestyle{myheadings}
\makeatletter\newcommand{\ps@icrc}{\renewcommand{\@oddhead}{\slshape{OG.2.2.05}\hfil}}
\makeatother\thispagestyle{icrc}

%\markright{OG.2.2.05}
%  Title:
\begin{center}
{\LARGE \bf Search for VHE pulsed emission from the Crab with the CAT Telescope}
\end{center}

%  Author List:
\begin{center}
{\bf A. Musqu\`ere$^{1}$, for the CAT collaboration$^{1,2,3}$}\\

{\it $^{1}$ CESR, CNRS/UPS, Toulouse, France (musquere@cesr.fr) \\
$^{2}$ IN2P3/CNRS and DAPNIA/CEA, France \\
$^{3}$ Charles University, Prague, Czech Republic}
\end{center}

%  Abstract:
\begin{center}
{\large \bf Abstract\\}
\end{center}
\vspace{-0.5ex}

Since 1996, the CAT experiment, operating at the THEMIS site (French
Pyrénées), has been collecting Very High Energy (VHE) $\gamma$-ray
data from the Crab. The temporal analysis of photon arrival times
folded with the pulsar parameters did not reveal any significant
pulsation. The upper limit of a steady pulsed flux over the 102.7
hours of observation is 1.5 10$^{-12}$ \Fpho, 3.0 10$^{-13}$
\Fpho~and 5.4 10$^{-14}$ \Fpho~above 250 GeV, 1 TeV and 5 TeV,
respectively. These results put stringent constraints on the models of
high energy pulsar electrodynamics.

\vspace{1ex}

\section{Introduction}
\label{intro.sec}

The Crab Nebula has an extraordinarily broad spectrum, mostly
attributed to synchrotron radiation from the relativistic wind of
electrons, accelerated up to a few TeV and injected by the pulsar in
the surrounding nebula (De Jager \& Harding 1992, Atoyan \& Aharonian
1996). However, above $\sim$ 1 GeV, a high energy spectral component
is thought to be produced by inverse Compton scattering between
``seed'' photons (essentially the synchrotron spectrum of the Nebula
in the Crab case) and the relativistic wind. This continuous emission
has been detected by many VHE groups (Weekes et al. 1989, Vacanti et
al. 1991, Akerlof et al. 1990, Baillon et al. 1991, Goret et al.,
1993, 1997, Iacoucci \& Nuss 1998) and its integral flux appears
constant making the Crab Nebula the standard candle in VHE astronomy.

Pulsed radiation has been detected up to GeV energies. Although some
groups have reported transient TeV pulsed emission, long term
observations of the pulsar show no modulation. In particular, a
stringent upper limit for a pulsed emission with 10 \% duty cyle of
$1.7\times 10^{-12}$ \Fpho~has been derived above 300 GeV by the
Whipple group (Gillanders et al. 1997).

In this paper, we present the results of the search for pulsed
emission from the Crab with the CAT telescope. First, we briefly
describe the instrument and the methods used for selecting VHE
photons. Then, upper limits are derived and compared to results
obtained with similar experiments. Finally, implications for the
physics of the pulsar and possible developments of this study are
briefly discussed.

\section{The CAT VHE telescope and the data reduction}
\label{cat.sec}

The CAT imaging telescope uses the Atmospheric Cherenkov Technique
(ACT) to observe Very High Energy $\gamma$-ray sources. A
comprehensive description of the experimental setup can be found in
Barrau et al. (1998). This setup features a very high definition
camera consisting of 600 phototubes (PMT) at the focus of a 17.8~m$^2$
mirror. The CAT experiment has been operating since September 1996 in the
French Pyrénées on the site of the former Themis solar power plant.
The trigger requires a signal greater than 3 photoelectrons in at
least 4 PMTs. With such conditions, the trigger rate is about $\sim$
15 Hz at zenith. When triggered, the camera records the image of
cosmic-ray showers through Cherenkov light. Each triggered event gets
a time stamp using a GPS (Global Positioning System) clock.

%Extensive Monte-Carlo
%simulations of gamma-ray Cherenkov showers have been performed to
%create a catalog of theoretical average images as a function of impact
%parameter and energy. An analytical model deduced from this catalog
%(Le Bohec et al. 1998) is fitted (using a ${\chi^2}$-like method) to
%the individual images to determine the characteristics of the incident
%$\gamma$-ray, i.e. the pointing angle ($\alpha$ : angle at the image
%barycenter between the actual source position and the reconstructed
%direction) and the energy.
Due to the fine grain of the CAT camera, a powerful image analysis is
performed, based on a maximum likelihood method using the longitudinal
and transverse distributions of Cherenkov light in each image (Le
Bohec et al. 1998).  It provides a $\chi^2$-like goodness-of-fit
parameter ($P_{\chi^2}$) allowing a good gamma-ray/hadron separation
as well as a measurement of the direction ($\alpha$) and energy of
electromagnetic shower.

The CAT standard selection of $\gamma$-ray-like events is made with
the following cuts : $P_{\chi^2}>0.35$, $\alpha < 6^\circ$, Q$_{\rm
tot} > 30$ where Q$_{\rm tot}$ is the total number of photoelectrons
in the image. Simulations show that this selection yields to
$\gamma$-ray efficiency better than $40\%$ and a rejection factor for
hadrons of $\sim 200$. For this particular combination of cuts, we
have determined (from the MC simulations) the effective area
(A$_{eff}$(E,$\theta$) in cm$^2$) as a function of energy and zenital
pointing angle ($\theta$).

\section{Search for pulsed emission}
\label{pulsed.sec}

Since september 1996, the Crab nebula has been observed with the CAT
telescope during ON scans for 102.7 hours. Arrival times of individual
events were converted from Universal Coordinated Time to Barycentric
Dynamical Time, then corrected for the propagation delay to the solar
system barycenter using the JPL DE200 planetary ephemeris (Standish
1982). Corresponding radio phases were obtained from the Jodrell Bank
ephemeris (Lyne 1999). To search for a possible modulation at the
pulsar frequency, we applied the standard selection criteria described
above and construct 60 bin phasograms with selected events for various
combinations of observing time and energy ranges. Each of these
phasograms were tested for non-uniformity using a Pearson's $\chi^2$
test. These phasograms show no evidence for any peculiar excess,
especially if we consider the relatively high number of trials.

To derive an upper limit for the pulsed emission, two assumptions must
be made : the pulsar duty cycle (10~\% centered on radio origin) and
the spectral shape (a power law with a differential spectral index
$\gamma$ = 2.15). Both assumptions are consistent with the EGRET
observations at lower energies (Nolan et al. 1993). Knowing the total
number of events above the cuts in a given energy range ($\Delta$E),
we can derive a 99.9 \% confidence level upper limit for the number of
pulsed photons (N$_{ul}$) over the full observations. This number is
converted to an integral flux upper limit (F$_{ul}$) by dividing its
value by the total acceptance of the telescope (A$_{tot}$, in cm$^2$~
sec). This acceptance is calculated by averaging the effective area
(A(E,$\theta$)) over the expected spectrum and summing the
contributions of all the runs we used. In these calculations, we take
into account the energy resolution of the telescope and the actual
zenith angle.

%\begin{center}
% $$\displaystyle A_{acc} = \frac
%{\int_{\Delta E} E^{-\gamma} \int_{Obs.} A_{eff}(E,\theta(t)) dt
%dE}{\int_{\Delta E} E^{-\gamma} dE}$$
%\end{center}

The results of this analysis are reported for three energy thresholds
in Table 1. The upper limits are then compared to previously published
ones in Figure 1. The CAT upper limits presented here are consistent
(even slightly better) with the upper limit reported by the Whipple
group above 300 GeV (Gillanders et al. 1997).

\begin{figure*}[!t]

\begin{center}

\psfig{figure=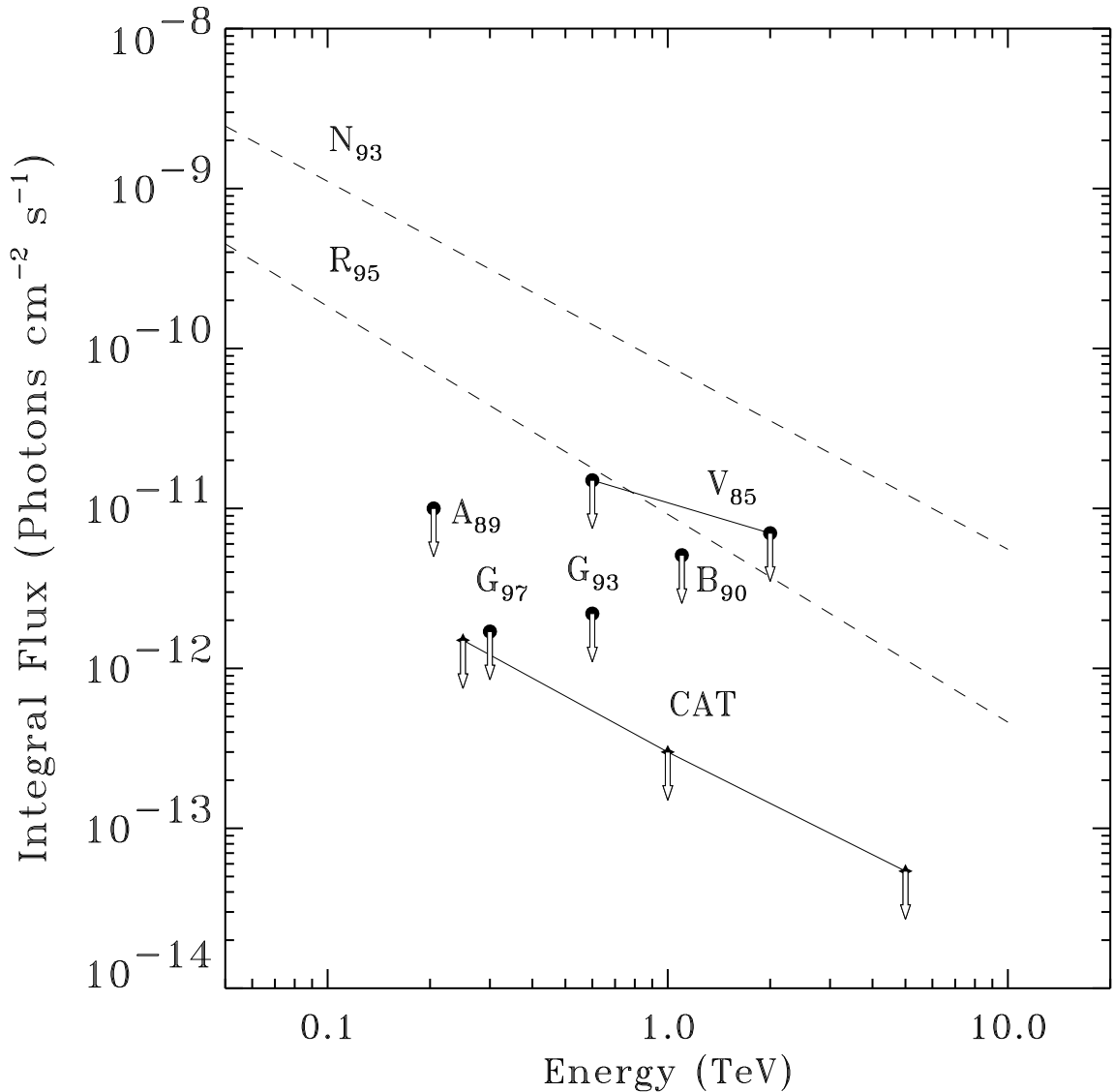,width=16.cm,angle=0}

\caption{Upper limits on the pulsed integral flux reported for
the Crab pulsar. B$_{90}$ : Bhat et al. 1990; V$_{85}$ : Vishwanath et
al. 1985; A$_{89}$ : Akerlof et al. 1989; G$_{93}$ : Goret et
al. 1993; G$_{97}$ : Gillanders et al. 1997. The dashed lines
represent the extrapolations of the EGRET integral pulsed
flux. N$_{93}$ : Nolan et al. 1993 (Full energy range); R$_{95}$ :
adapted from Ramanamurthy et al. 1995 (above 1 GeV)}

\end{center}

\end{figure*}

\begin{table}[!h]
\begin{center}
\begin{tabular}{|c|c|c|c|c|}
\hline

 $\Delta E$ & N$_{tot}$ &  N$_{ul}$ & A$_{tot}$         &  F$_{ul}$   \\
 (GeV)      &           &           & (cm$^{2}$ sec)      &  (ph cm$^{-2}$ sec$^{-1}$) \\
\hline
\hline

$ \geq $ 250   &  28342  &  168   &  1.1 10$^{14}$     &  1.5 10$^{-12}$  \\
\hline
$ \geq $ 1000  &  10251  &  102   &  3.4 10$^{14}$     &  3.0 10$^{-13}$   \\
\hline
$ \geq $ 5000  &  679    &  27    &  5.0 10$^{14} $    &  5.4 10$^{-14}$   \\
\hline

\hline
\hline

\end{tabular}

\caption{Results of the pulsed analysis for three energy thresholds.
N$_{tot}$ is the total number of events above the standard cut,
N$_{ul}$ is the 99.9 \% confidence level upper limit on the number
of pulsed photons (10 \% duty cycle), A$_{acc}$ is the total
acceptance of the telescope (see text), and F$_{ul}$ the
integral flux upper limit.}

\end{center}
\end{table}

\section{Discussion}
\label{discussion.sec}

Pulsed emission from the Crab, characterized by a power-law spectrum,
is possibly generated in the corotating magnetosphere by relativistic
particles accelerated in vacuum gaps. In the outer gap model, primary
charged particles accelerated to highly relativistic energies emit
curvature and/or inverse-compton radiation. A synchrotron emission
(from optical to $\gamma$-rays) is produced by the secondary particles
generated in the induced electromagnetic cascades (Cheng, Ho \&
Ruderman 1986 see also Romani 1996 and references therein). A revised
version of this model (Romani 1996, Romani \& Yadigaroglu 1995),
including the Compton upscattering of the synchrotron emission on the
primary e$^{+/-}$ predicts (for a standard $\gamma$-ray pulsar : B=3
10$^{12}$ Gauss, synchrotron cutoff at 3 GeV) a TeV {\it pulsed}
emission containing $\sim$ 1 \% of the GeV flux and showing up in the
spectrum well above the synchrotron cutoff.

In the polar cap model, the acceleration region is located above the
neutron star polar caps, in a region of a few NS radii which encloses
open magnetic field lines (Daugherty \& Harding 1996, Harding \&
Muslimov, 1998 and references therein). In this case, primary
particles of maximum energy between $\sim$ 5 and 50 TeV are produced
across the gap. These high energy particles radiate a VHE emission via
curvature radiation and/or inverse Compton scattering with seed
photons (possibly the soft thermal emission coming from the neutron
star surface). The escape probability of the produced photons
crucially depends on attenuation processes such as the magnetic pair
creation close to the neutron star.

The EGRET observations of the Crab suggest that a spectral cut-off or
a break occurs around 1 GeV with a integral spectral index of -0.88
below this value and -1.3 above it (Ramanamurthy et al. 1995). The
upper limits obtained with the CAT experiment represent $\sim$ 3 \% of
the extrapolation of the EGRET spectrum (see Figure 1, R$_{95}$)
strongly suggesting that spectrum above few GeV is much steeper than
predicted with a simple broken powerlaw. On the other hand, our upper
limits give a stringent constraint for a TeV pulsed component such as
the one discussed above.

The above discussion demonstrates that the quest for a pulsed emission
from the Crab pulsar above a few GeV must go on. First, more
observations have to be performed with imaging VHE telescopes and the
Crab definitively remains a major target for CAT. Second, the analysis
scheme might also be extended to steeper pulsar spectra with different
cuts adapted to the corresponding spectral index. This study will be
reported in a future work. Finally, future observations in the 1-200
GeV energy range with instruments such as CELESTE, HESS and GLAST will
certainly provide new insights.

\vspace{1ex}

\begin{center}

{\Large\bf References}

\end{center}

\noindent Akerlof C.W., et al. 1990, Proc. of the XXI ICRC (Adelaide), {\bf 2}, 135 \\
Akerlof C.W., et al. 1989,  Proc. of the GRO Workshop (Greenbelt, NRL), 4-49 \\
Atoyan A.M., \& Aharonian, F.A. 1996, MNRAS, {\bf 278}, 525 \\
Baillon P., et al. 1991, Proc. of the XXII ICRC (Dublin), {\bf 1}, 220 \\
Barrau, A., et al. 1998, NIM A, {\bf 416}, 278\\
Bhat P.N., et al. 1990, Proc. of the XXI ICRC (Adelaide), {\bf 2}, 148\\
Chen K.Y., Ho C. \& Ruderman M.A., 1986,  Astrophys. J., {\bf 300}, 522 \\
Daugherty J.K., \& Harding A.K., 1996, Astrophys. J., {\bf 458}, 278 \\
de Jager O.C., \& Harding, A.K. 1992, Astrophys. J., {\bf 396}, 161\\
Gillanders G.H., et al. 1997 , Proc. of the XXV ICRC (Durban), {\bf 3}, 185\\
Goret P., et al. 1993, A\&A, {\bf 270}, 401 \\
Goret P., et al. 1997, Proc. of the XXV ICRC (Durban), {\bf 3}, 173 \\
Harding A.K. \& Muslimov A.G., 1998, Astrophys. J., {\bf 508}, 328 \\
Iacoucci L., \& Nuss E., 1998,  Proc. of the XVI ECRS (Madrid), {\bf G.R.2.5}, 363\\
Le Bohec S., et al. 1998, NIM A, {\bf 416}, 425 \\
Lyne A.G., 1999, Jodrell Bank Crab Monthly Ephemeris
(http://www.jb.man.ac.uk/$\sim$pulsar/crab.html) \\
Nolan P.L., et al. 1993, Astrophys. J., {\bf 409}, 697 \\
Ramanamurthy P.V., et al. 1995, Astrophys. J., {\bf 450}, 791 \\
Romani R.W., 1996,  Astrophys. J., {\bf 470}, 469 \\
Romani R.W. \& Yadigaroglu I.A., 1995  Astrophys. J., {\bf 438}, 314 \\
Standish M., 1982, A\&A, {\bf 114}, 297 \\
Vacanti G., et al. 1991, Astrophys. J., {\bf 377}, 467 \\
Vishwanath P.R., et al. 1985,  Proc. of the IXX ICRC (La Jolla), {\bf 1}, 144\\
Weekes T.C., et al. 1989, Astrophys. J., {\bf 342}, 379\\

\end{document}